\documentclass[letterpaper, 10 pt, conference]{ieeeconf}  

\IEEEoverridecommandlockouts                              

\usepackage{graphics} 
\usepackage{epsfig} 
\usepackage{mathptmx} 
\usepackage{times} 
\usepackage{amsmath} 
\usepackage{amssymb}  
\usepackage{hyperref}
\usepackage{cleveref}
\usepackage{todonotes}
\usepackage{multirow}

\let\labelindent\relax
\usepackage{enumitem}

\usepackage{array}
\newcolumntype{P}[1]{>{\centering\arraybackslash}p{#1}}

\usepackage{xcolor}

\AtBeginDocument{}

\newcommand\copyrighttext{%
  \footnotesize \textcopyright 2024 IEEE.  Personal use of this material is permitted.  Permission from IEEE must be obtained for all other uses, in any current or future media, including reprinting/republishing this material for advertising or promotional purposes, creating new collective works, for resale or redistribution to servers or lists, or reuse of any copyrighted component of this work in other works.}
\newcommand\copyrightnotice{%
\begin{tikzpicture}[remember picture,overlay]
\node[anchor=south,yshift=10pt] at (current page.south) {\fbox{\parbox{\dimexpr\textwidth-\fboxsep-\fboxrule\relax}{\copyrighttext}}};
\end{tikzpicture}%
}

\title{\LARGE \bf
Precise Workcell Sketching from Point Clouds Using an AR Toolbox
}

\author{Krzysztof Zieli\'nski$^{1,2}$, Bruce Blumberg$^{2}$, Mikkel Baun Kjærgaard$^{1}$
\thanks{$^{1}$Authors are with
the Faculty of Engineering, the Maersk Mc Kinney Moller Institute, University of Southern Denmark
        {\tt\small krzi, mbkj@mmmi.sdu.dk }}%
\thanks{$^{2}$Authors are with Universal Robots A/S
        {\tt\small krzi, brbl@universal-robots.com}}%
}

\begin{document}

\maketitle
\copyrightnotice
\thispagestyle{empty}
\pagestyle{empty}

\begin{abstract}
Capturing real-world 3D spaces as point clouds is efficient and descriptive, but it comes with sensor errors and lacks object parametrization. These limitations render point clouds unsuitable for various real-world applications, such as robot programming, without extensive post-processing (e.g., outlier removal, semantic segmentation). On the other hand, CAD modeling provides high-quality, parametric representations of 3D space with embedded semantic data, but requires manual component creation that is time-consuming and costly. To address these challenges, we propose a novel solution that combines the strengths of both approaches. Our method for 3D workcell sketching from point clouds allows users to refine raw point clouds using an Augmented Reality (AR) interface that leverages their knowledge and the real-world 3D environment. By utilizing a toolbox and an AR-enabled pointing device, users can enhance point cloud accuracy based on the device’s position in 3D space. We validate our approach by comparing it with ground truth models, demonstrating that it achieves a mean error within 1cm --- significant improvement over standard LiDAR scanner apps.

\end{abstract}

\section{Introduction}

Large manufacturing companies have used robot automation for decades to streamline production processes. Advancements in flexible manufacturing systems have made robots (such as collaborative robots) more accessible and cost-effective for Small and Medium Enterprises (SMEs). Despite this progress, deploying a robotic system remains costly and requires specialized expertise. Creating a digital twin~\cite{rossmann_new_2013} to validate system requirements involves modeling the workcell in simulation software, which can be time-consuming and expensive. Simplifying this step is crucial to encourage companies to automate their production lines.

Currently, simulation software that is used to create digital twins requires CAD models or simplified 3D primitives of the objects used in the workcell. While 3D models offer valuable parametrization capabilities, allowing easy scaling, movement, addition, or removal of objects within the workcell, they come with certain limitations. Two widely used robot simulation software tools in the industry, ABB RobotStudio~\cite{website:RobotStudio} and RoboDK~\cite{website:RoboDK}, exemplify this trade-off. These tools excel in accuracy and flexibility, but at a cost: high entry-level knowledge requirements and significant time consumption. Moreover, when dealing with complex shapes, additional modeling tools may be necessary before importing the models into the aforementioned software.

An alternative to professional modeling tools is point cloud capture and representation, available as standard mobile phone apps. It is a very fast method to capture a scan of the real-world workcell, ensuring completeness of the capture without the need to manually get measurements of the workcell. Previous work has explored annotating such point clouds to use them for helping with robot deployment~\cite{zielinski_robotgraffiti_2024}. While point clouds offer advantages, such work has also identified limitations. Point clouds lack parametric properties and semantic object data, making post-capture alterations challenging. Digital twins used for simulation often require additional objects not present during scanning (e.g., robot stand) or exclude elements captured during scanning but unnecessary for the final workcell (e.g., operator station). Moreover, the sensors used for point cloud capture, such as LiDAR, have inherent limitations. Dark surfaces absorb light, creating holes in the scan; shiny surfaces cause uncontrollable light reflections, resulting in outliers (and robots usually have metallic components); clear surfaces let the light go through (and workcells are often isolated behind transparent walls). Unfortunately, consumer-grade LiDAR scanners lack the necessary precision, while professional-grade scanners are prohibitively expensive for many applications. Based on these limitations, point clouds are either not used for robot deployment or require a lot of post-processing workflows in order to clean and improve the accuracy of the scan and optionally convert them into parametric models, which is unpractical from an SME perspective.

\begin{figure}[h]
    \begin{center}
        \includegraphics[width=\linewidth]{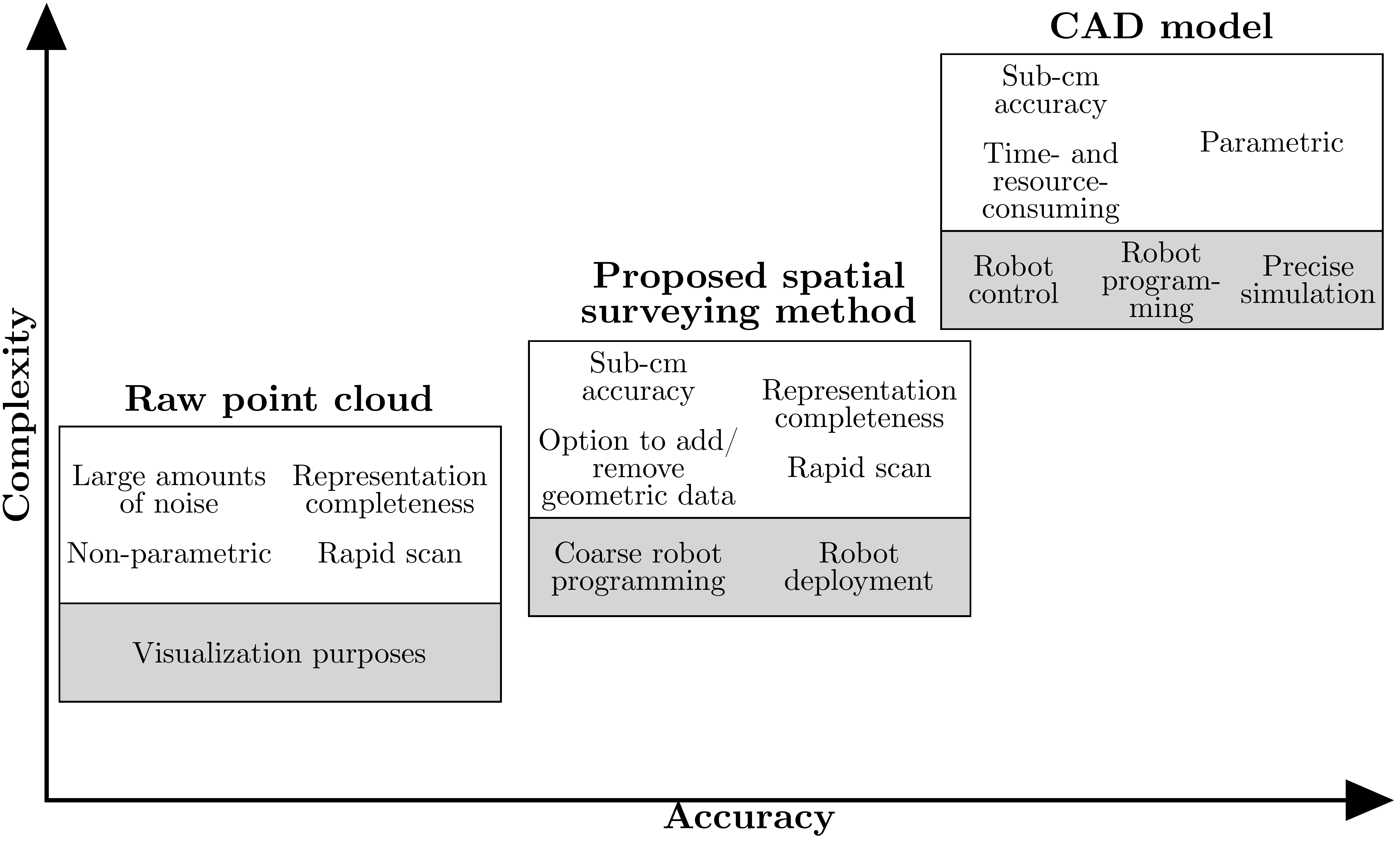}
        \caption{Digital representation types placed based on complexity and accuracy levels. \textit{Top}: representation characteristics, \textit{Bottom}: robot applications.}
        \label{fig:ScanningMethods}
    \end{center}
\end{figure}

We propose a novel method for 3D workcell sketching from point clouds, that bridges the gap between fast but imprecise point clouds and meticulous but time-consuming CAD models --- illustrated in \autoref{fig:ScanningMethods}. Our method starts with a point cloud representation enhanced with an AR-enabled toolbox. This toolbox empowers users to process the point cloud on the collection device itself, leveraging their knowledge and real-world context. It allows users to improve the accuracy of the scan, remove unnecessary points, and add basic 3D primitives that might not exist in the physical workcell.

The main contributions of this paper are the method for 3D workcell sketching from point clouds, enabled by the novel AR-enabled User Interface for on-device point cloud processing, implementation of the proposed method, and a comparative study on the accuracy of the proposed solution and raw point cloud scan.

In \cref{sec:related_work}, we describe existing work within point cloud processing and 3D sketching with a focus on AR and Virtual Reality (VR). \Cref{sec:method} describes the proposed method, with results and discussion of the comparative study in \cref{sec:results} and \autoref{sec:discussion}.

\section{Related work}
\label{sec:related_work}

In this work, we investigate existing point cloud processing and selection techniques within augmented and virtual reality environments. Subsequently, we draw upon 3D sketching methodologies as a source of inspiration for enhancing user interaction within augmented reality. Finally, we examine the latest advancements in point cloud processing methods.

\subsection{Point cloud processing in AR and VR}
Point clouds represent 3D data, inherently challenging to manipulate using 2D displays and cursors. Consequently, AR and VR have emerged as promising solutions for point cloud interaction. Most research focuses on point cloud annotation within 3D space. \textit{Go'Then'Tag}~\cite{veit_gothentag_2014} employs a phone as a 3D input device (pencil-like) to select points in a VR environment. Lin et al.~\cite{lin_immersive_2023} use cube primitives for selecting and labeling points within their extent. Additionally, Wirth et al.~\cite{wirth_pointatme_2019} and Franzluebbers et al.~\cite{franzluebbers_virtual_2022} use VR controllers to create boxes and spheres, respectively, around desired points for machine learning labeling purposes. Alternatively, \textit{Touching the Cloud}~\cite{lubos_touching_2014} and \textit{Slice-n-Swipe}~\cite{bacim_slice-n-swipe_2014} methods track users hands in 3D space for point selection.
When dealing with large point clouds, rendering strategies become crucial to accommodate devices with limited computational capabilities. Techniques like hierarchical data structures, as suggested by Casado-Coscolla et al.~\cite{casado-coscolla_rendering_2023}, provide continuous levels of detail for point clouds. Additionally, Fan et al.~\cite{fan_pcqdar_2024} explore users’ perceived quality of presented point clouds in AR.

\subsection{3D sketching in AR}
3D sketching is a narrow field of user interaction within AR where the user is given tools to interact within the real/virtual world. It fits well with point cloud selection methods. \textit{Mobi3DSketch}~\cite{kwan_mobi3dsketch_2019} is a method for drawing 3D sketches on a 2D screen with AR-enabled device. Similarly, Börsting et al.~\cite{borsting_design_2022} create a virtual spray can for drawing in 3D space within a real environment. Mi et al.~\cite{mi_mixed_2021} use a 3D-tracked marker to draw shared 3D space with other users. Wilkes et al.~\cite{wilkes_3d_2012} use a head-mounted AR/VR display and a phone as a tracking and drawing device. Barber et al.~\cite{barber_sketch-based_2010} project 2D sketches into the 3D environment for control of a small mobile robot.

\subsection{Point cloud processing methods}
Conventional point cloud processing methods are well-known and implemented in widely used open-source libraries --- Open3D~\cite{open3D} and PCL~\cite{Rusu_ICRA2011_PCL}. Most of the current research involves deep neural networks (DNNs). \textit{PointCleanNet}~\cite{rakotosaona_span_2020} is a deep-learning-based de-noising and outlier removal method, with Roman-Rivera et al.~\cite{mata-rivera_3d_2023} building on top of it. Wang et al.~\cite{wang_lightn_2024} and Xiong et al.~\cite{xiong_semantic_2023} use DNNs for point cloud downsampling preserving geometric and semantic properties, respectively.

We draw inspiration for our method from 3D sketching tools in AR, leverage the advantages of 3D space for point cloud selection, and implement state-of-the-art methods for on-device point cloud processing.

\section{Method}
\label{sec:method}

The method for 3D workcell sketching from point clouds is realized as a mobile application for on-device point cloud capture using LiDAR and point cloud processing. It is designed for a novice user wanting to evaluate the automation feasibility of their production line with available tools, such as smartphones. It results in cleaned-up point cloud representation optionally enhanced with virtual objects. The obtained point cloud can be then used for further processing, such as evaluating workcell metrics for robot automation in simulation software.

\begin{figure*}[h]
    \begin{center}
        \includegraphics[width=\textwidth]{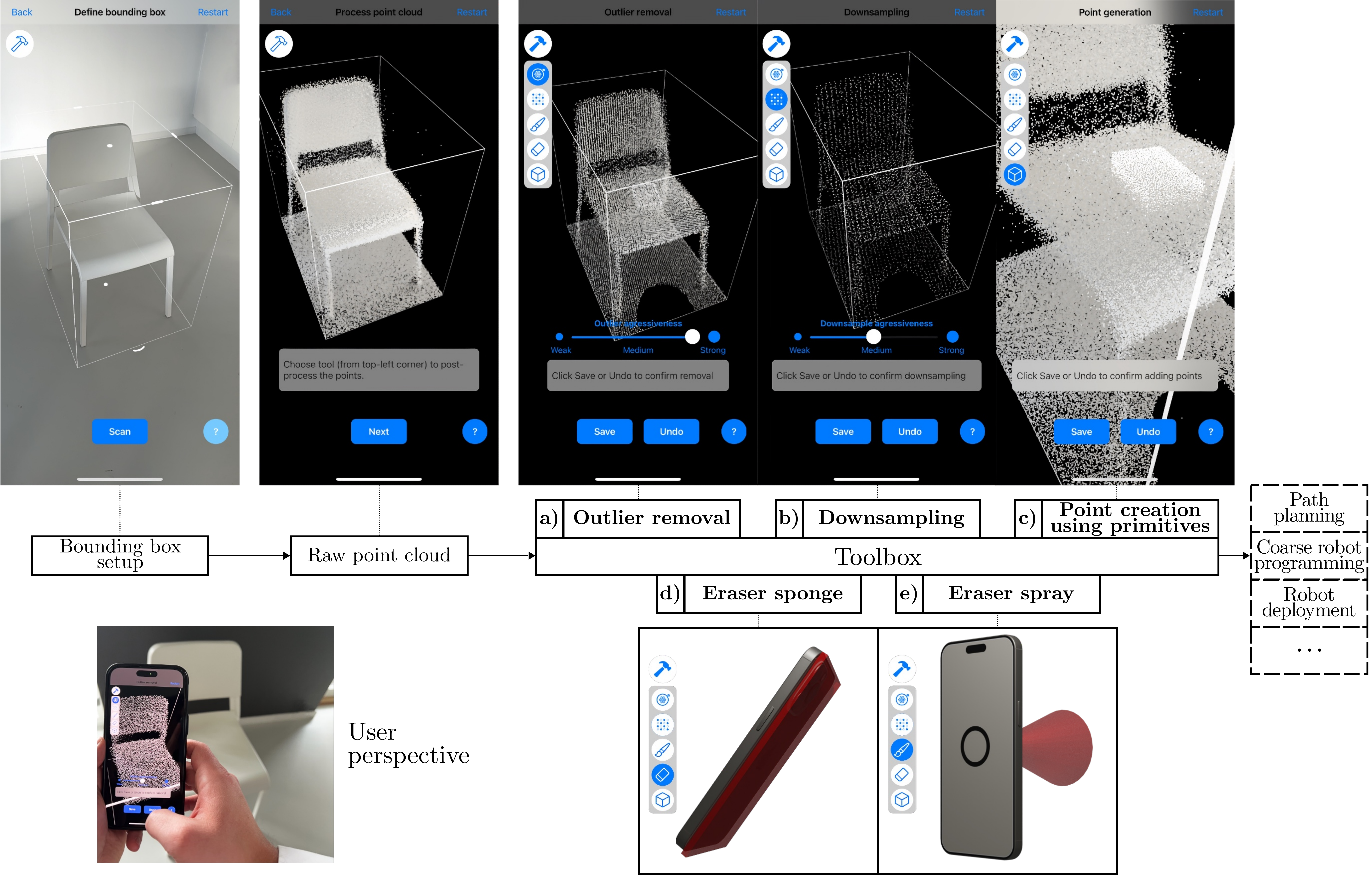}
        \caption{The flow of the point cloud collection with each tool available in the toolbox: a) outlier removal, b) downsampling, c) point creation using primitives, d) eraser sponge, and e) eraser spray.}
        \label{fig:MethodFlow}
    \end{center}
\end{figure*}

First, we describe the flow of point cloud collection and then, available point cloud processing tools. The flow and the available tools in the toolbox are presented in \autoref{fig:MethodFlow}.

\subsection{Point cloud collection flow}
The user is presented with a camera view of the workcell. The first step is to set up a bounding box around the area to be captured to limit unnecessary point collection (background noise) outside of the area of interest. The user is suggested with a box based on feature points (distinctive points within the view) that automatically resizes when the device is moved around. The user can fine-tune the dimensions and position of the box using graspable handles. Then, the user is asked to walk around and capture the points within the chosen bounding box. We capture RGB data in a grid that is enhanced with depth data from a LiDAR sensor, consequently projected in a 3D space.

\subsection{Point cloud processing tools}
We have created tools in the app that allow users to enhance the captured point cloud. For every tool, we present the user with a preview that gives the possibility to be saved or discarded. This toolbox is designed specifically for interaction with point clouds. We present each tool below:
\begin{enumerate}[label=\autoref{fig:MethodFlow}\alph*) , wide=0pt, font=\itshape]
    \item \textbf{Outlier removal} --- used to clean noise from the scan, such as incorrectly projected points from a reflective surface. The user can choose how aggressive the outlier removal should be (\textit{weak, medium, strong}) which corresponds to the standard deviation threshold of the average distances across the point cloud. We use statistical outlier removal that removes points furthest away from their neighbors.
    \item \textbf{Downsampling} --- used to reduce point density, remove overlapping points, and create an organized structure. The user can choose the aggressiveness of downsampling --- \textit{weak, medium, strong}, that corresponds to the voxel size that collects all points into a box and replaces them with one point per voxel.
    \item \textbf{Point creation using primitives} --- used when the real workcell is missing objects that are supposed to be in the final workcell. The tool allows the user to place a rectangular prism on any vertical or horizontal surface of the physical workcell. The user can then modify the size of the primitive to the needs, and once set, the object is sampled and added to the point cloud representation.
    \item \textbf{Eraser sponge} --- a precise tool for point removal in difficult spaces, such as inside a CNC machine. The mobile device acts like a virtual blackboard sponge where the user moves the device in physical space to remove points in the point cloud scan. This is accomplished with on-device tracking (visual and inertial odometry) and a rectangular prism around the device that removes points if within its bounds. The user can choose how thick the sponge should be --- \textit{small, medium, big}.
    \item \textbf{Eraser spray} --- an alternative tool for point removal for less precise needs, such as removal of captured walls or floors adjacent to the captured object. To use this tool, the user can tap and drag a finger on the screen to create a virtual spray in the shape of a flipped cone. Any point that is within the cone is removed from the point cloud. This tool provides 2 modifiers: spray size --- \textit{small, medium, big} used for the radius of the cone and spray depth --- \textit{shallow, medium, deep} for the height of the cone.
\end{enumerate}

The above-mentioned tools offer cloud-wide processing --- \autoref{fig:MethodFlow}a--b, and point-specific processing --- \autoref{fig:MethodFlow}c--e, providing a wide scope of point cloud processing needs. In \cref{sec:results}, we evaluate these tools.

\section{Evaluation and Results}
\label{sec:results}

We conduct a comparative study to evaluate the improvement in accuracy of the point cloud processed with our method for 3D workcell sketching compared to the method of collecting unprocessed raw point cloud. Moreover, we designed a set of robot automation workcells to exemplify different robot deployment scenarios that this tool might be useful for.

\subsection{Evaluation scenarios}
\begin{figure*}[h]
    \begin{center}
        \includegraphics[width=\textwidth]{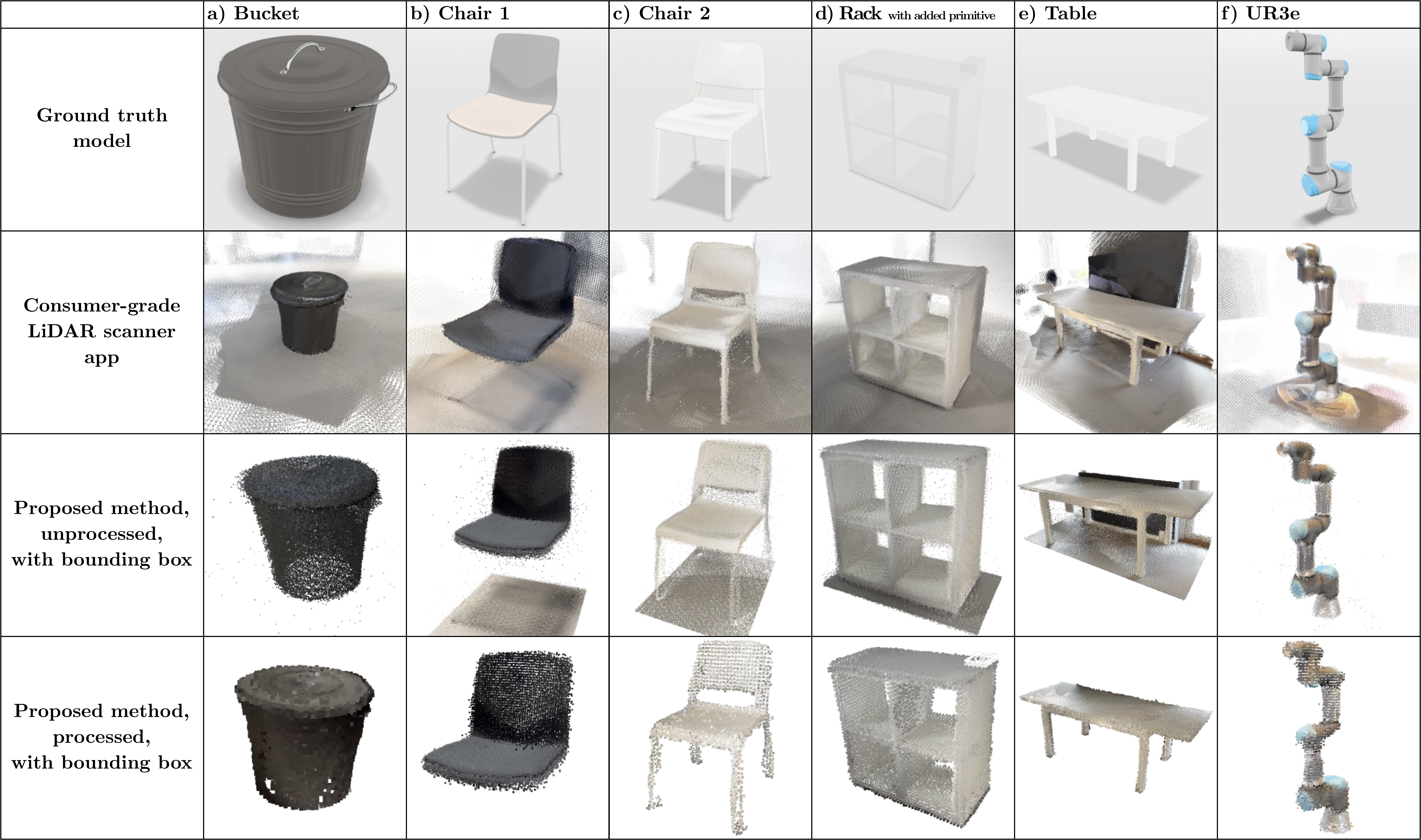}
        \caption{3D models that were used to obtain evaluation metrics in \autoref{tab:error_3DModels}. Objects sorted based on test scenarios: \ref{sce:1} shiny surface difficult to model --- (a,b,f); \ref{sce:2} scans containing unnecessary data --- (e); \ref{sce:3} scans containing background noise, e.g., floors --- (a--f); \ref{sce:4} --- objects missing geometrical data at the time of the scan (d).}
        \label{fig:3Dmodels}
    \end{center}
\end{figure*}

To evaluate the overall functionality of the proposed method, we have prepared scenarios that the user might stumble upon when capturing real-world workcell. These scenarios have been selected to cover the limitations of the LiDAR scanners and workcell constraints. The objects that have been used for this comparative study also reflect those possible cases and are presented in \autoref{fig:3Dmodels}. The scenarios are the following:

\begin{enumerate}[label=Scenario \arabic*. , wide, font=\itshape]
\label{enum:scenarios}
    \item an object that contains difficult-to-capture surfaces, i.e., dark --- consumes light, shiny --- bounces off light, clear --- light passes through. \autoref{fig:3Dmodels}a,b,f) --- metallic bucket handle, metallic chair legs, metallic UR3e tubes: due to shiny surfaces, scans contain incorrectly projected outliers.\label{sce:1}
    \item an object adjacent to another object that is not supposed to be in the digital representation. \autoref{fig:3Dmodels}e) --- a table standing against a wall that is not an intended part of the workcell. \label{sce:2}
    \item an object that contains background noise. \autoref{fig:3Dmodels}a--f) --- all the scanned objects contain background noise, such as the floor. \label{sce:3}
    \item an object missing parts that are supposed to be in the digital representation. \autoref{fig:3Dmodels}d) --- a rack that is supposed to have a box placed on top of it, but it was missing during the scan. \label{sce:4}
\end{enumerate}


\subsection{Proposed method implementation}
To evaluate our method we implement it as a mobile application for iOS devices in Swift. For a fast and smooth user experience, we perform point cloud visualization from raw data on the device's GPU using MetalKit~\cite{website:MetalKit}. To perform point cloud capture, for every frame, we capture an RGB image and depth map. We define grid size (2000 points per frame yielding $8.7mm$ resolution) that is used to sample RGB and depth values from the image plane. Lastly, the RGB-D point is projected to 3D world space using the camera's projection transform. Every time, the device either moves by $0.01 m$ or rotates by $1^{\circ}$, new capture is run and new RGB-D points are added to the scan. Using Metal shaders, we calculate whether points should be displayed or not (when outside the bounding box or erased using eraser spray or sponge). For AR experiences, such as placing a bounding box or primitive on flat surfaces, we use ARKit~\cite{website:ARKit}. For processing of outlier removal and downsampling, we use Open3D library~\cite{open3D}.

\subsection{Evaluation setup}
We evaluate our proposed method with the above-mentioned test scenarios. For comparison, we use an off-the-shelf consumer-grade LiDAR scanning app available on iOS --- SiteScape~\cite{website:SiteScape}, where previous work has conducted an accuracy comparison with professional Terrestrial Laser Scanner~\cite{spreafico_ipad_2021}. We also consider two versions of our proposed method: 1) using only an initial bounding box that allows us to limit the extent of the scan; and 2) using a bounding box and as many of the available tools within the toolbox as relevant. For all the methods, we collect a point cloud scan by walking around the object and exporting it to a PLY file. All the scans are collected using built-in LiDAR on an iPhone 15 Pro Max. The authors acted as users and thereby the evaluation results are for expert users. In future work, we plan to evaluate the input of different user skill levels.

\subsection{Evaluation metrics}
\begin{figure}[h]
    \begin{center}
        \includegraphics[width=\linewidth]{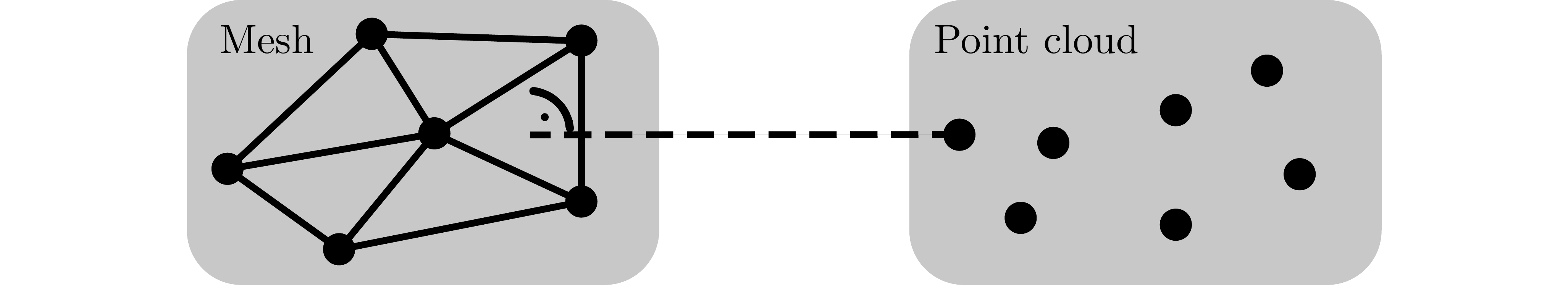}
        \caption{Cloud/mesh distance calculation.}
        \label{fig:DistanceError}
    \end{center}
\end{figure}

We evaluate the accuracy of the generated point clouds by comparing them with ground truth models. For evaluation, we use the \textit{CloudCompare} software~\cite{website:CloudCompare} with the following approach:
\begin{enumerate}[wide]
    \item load ground truth model of the object in \textit{OBJ} format and point cloud representation in \textit{PLY} format.
    \item approximately align both models by manually selecting point correspondences.
    \item sample the ground truth model with 10 million points for fine registration of the point cloud to the sampled ground truth model using Iterative Closest Point (ICP).
    \item use \textit{CloudCompare} cloud/mesh distance metric that finds the perpendicular distance between the point of the point cloud and closest triangle from the ground truth mesh, as seen in \autoref{fig:DistanceError}.
\end{enumerate}
The result is a point cloud color-coded as a heat map, that we export and process to obtain the following absolute distance metrics: mean, median, standard deviation, minimum, and maximum. We perform this evaluation based on the flow from~\cite{handa_benchmark_2014}. The absolute distance error of the captured objects visible in \autoref{fig:3Dmodels}, can be seen in \autoref{tab:error_3DModels}.


\begin{table}[]
\caption{Absolute distance error between captured point cloud and ground truth model. We compare a consumer-grade LiDAR scanner app; proposed method \textbf{(PM)}, unprocessed with bounding box; and the proposed method, processed with bounding box. Error in cm.}
\label{tab:error_3DModels}
\centering
\begin{tabular}{|c|l|l|l|l|}
\hline
\textbf{Object}          & \multicolumn{1}{c|}{\textbf{Error {[}cm{]}}} & \multicolumn{1}{c|}{\textbf{\begin{tabular}[c]{@{}c@{}}LiDAR\\ Scanner App\end{tabular}}} & \multicolumn{1}{c|}{\textbf{\begin{tabular}[c]{@{}c@{}}PM\\ unprocessed\end{tabular}}} & \multicolumn{1}{c|}{\textbf{\begin{tabular}[c]{@{}c@{}}PM,\\ processed\end{tabular}}} \\ \hline
\multirow{5}{*}{Bucket}  & Mean                                         & 32,130                                                                                    & 1,009                                                                                & \textbf{0,683}                                                                          \\
                         & Median                                       & 13,863                                                                                    & 0,532                                                                                & 0,535                                                                                   \\
                         & Std                                          & 45,648                                                                                    & 1,196                                                                                & \textbf{0,576}                                                                          \\
                         & Min                                          & 0,000                                                                                     & 0,000                                                                                & 0,000                                                                                   \\
                         & Max                                          & 414,258                                                                                   & 20,681                                                                               & 3,143                                                                                   \\ \hline
\multirow{5}{*}{Chair 1} & Mean                                         & 41,604                                                                                    & 8,740                                                                                & \textbf{0,849}                                                                          \\
                         & Median                                       & 22,480                                                                                    & 4,618                                                                                & 0,519                                                                                   \\
                         & Std                                          & 55,268                                                                                    & 8,565                                                                                & \textbf{0,878}                                                                          \\
                         & Min                                          & 0,000                                                                                     & 0,000                                                                                & 0,000                                                                                   \\
                         & Max                                          & 369,497                                                                                   & 41,520                                                                               & 6,569                                                                                   \\ \hline
\multirow{5}{*}{Chair 2} & Mean                                         & 37,195                                                                                    & 4,108                                                                                & \textbf{1,004}                                                                          \\
                         & Median                                       & 20,855                                                                                    & 0,469                                                                                & 0,812                                                                                   \\
                         & Std                                          & 46,753                                                                                    & 7,280                                                                                & \textbf{0,873}                                                                          \\
                         & Min                                          & 0,000                                                                                     & 0,000                                                                                & 0,001                                                                                   \\
                         & Max                                          & 438,837                                                                                   & 32,785                                                                               & 5,744                                                                                   \\ \hline
\multirow{5}{*}{Rack}    & Mean                                         & 35,119                                                                                    & 1,139                                                                                & \textbf{0,947}                                                                          \\
                         & Median                                       & 11,751                                                                                    & 0,735                                                                                & 0,818                                                                                   \\
                         & Std                                          & 63,855                                                                                    & 1,515                                                                                & \textbf{0,674}                                                                          \\
                         & Min                                          & 0,000                                                                                     & 0,000                                                                                & 0,000                                                                                   \\
                         & Max                                          & 654,049                                                                                   & 15,495                                                                               & 4,351                                                                                   \\ \hline
\multirow{5}{*}{Table}   & Mean                                         & 22,763                                                                                    & 13,790                                                                               & \textbf{1,121}                                                                          \\
                         & Median                                       & 14,639                                                                                    & 3,419                                                                                & 0,959                                                                                   \\
                         & Std                                          & 32,330                                                                                    & 16,608                                                                               & \textbf{0,990}                                                                          \\
                         & Min                                          & 0,000                                                                                     & 0,000                                                                                & 0,000                                                                                   \\
                         & Max                                          & 2857,806                                                                                  & 71,160                                                                               & 26,131                                                                                  \\ \hline
\multirow{5}{*}{UR3e}    & Mean                                         & 68,883                                                                                    & 0,854                                                                                & \textbf{0,796}                                                                          \\
                         & Median                                       & 25,732                                                                                    & 0,622                                                                                & 0,728                                                                                   \\
                         & Std                                          & 69,494                                                                                    & 0,808                                                                                & \textbf{0,630}                                                                          \\
                         & Min                                          & 0,000                                                                                     & 0,000                                                                                & 0,000                                                                                   \\
                         & Max                                          & 300,271                                                                                   & 29,436                                                                               & 3,787                                                                                   \\ \hline
\end{tabular}
\end{table}

\subsection{Evaluation metrics results}
For all the scanned objects, we have obtained smaller means and standard deviations of absolute distance error using our proposed method with processed points. Moreover, we have also obtained reduced means and standard deviations for our proposed method, only using a bounding box compared to a consumer-grade LiDAR scanner app. We see up to $86\times$ smaller mean value for \textit{UR3e} model between the off-the-shelf scanning app and our proposed method with processing. We also see up to $10\times$ reduction in the mean value for \textit{Chair 1} model between unprocessed and processed point clouds using our proposed method.

For all the scanned objects, our method has resulted mean absolute distance error within $1cm$, hence it shows significant improvements in the quality of the obtained point clouds. It is, however, comparing absolute distance error of scans that have not been processed prior at all, which is rarely the case. For most applications, point cloud representation has to be post-processed in order to obtain usable data. Our method proposes a simpler, on-device user interface that reaps on the benefits of AR and mobile touch interface. It is more intuitive to move in real-time and the real world with the device that acts as a virtual pointer in point cloud representation.

An important remark while evaluating absolute distance error is that it uses all the points in the point cloud scan for estimating the error, therefore it is natural that using customer-grade LiDAR scanner app and the proposed method with only a bounding box generate worse results because points that are not part of the actual model are taken into calculation.

\subsection{Examples of robot automation workcells}
\begin{figure}[h]
    \begin{center}
        \includegraphics[width=\linewidth]{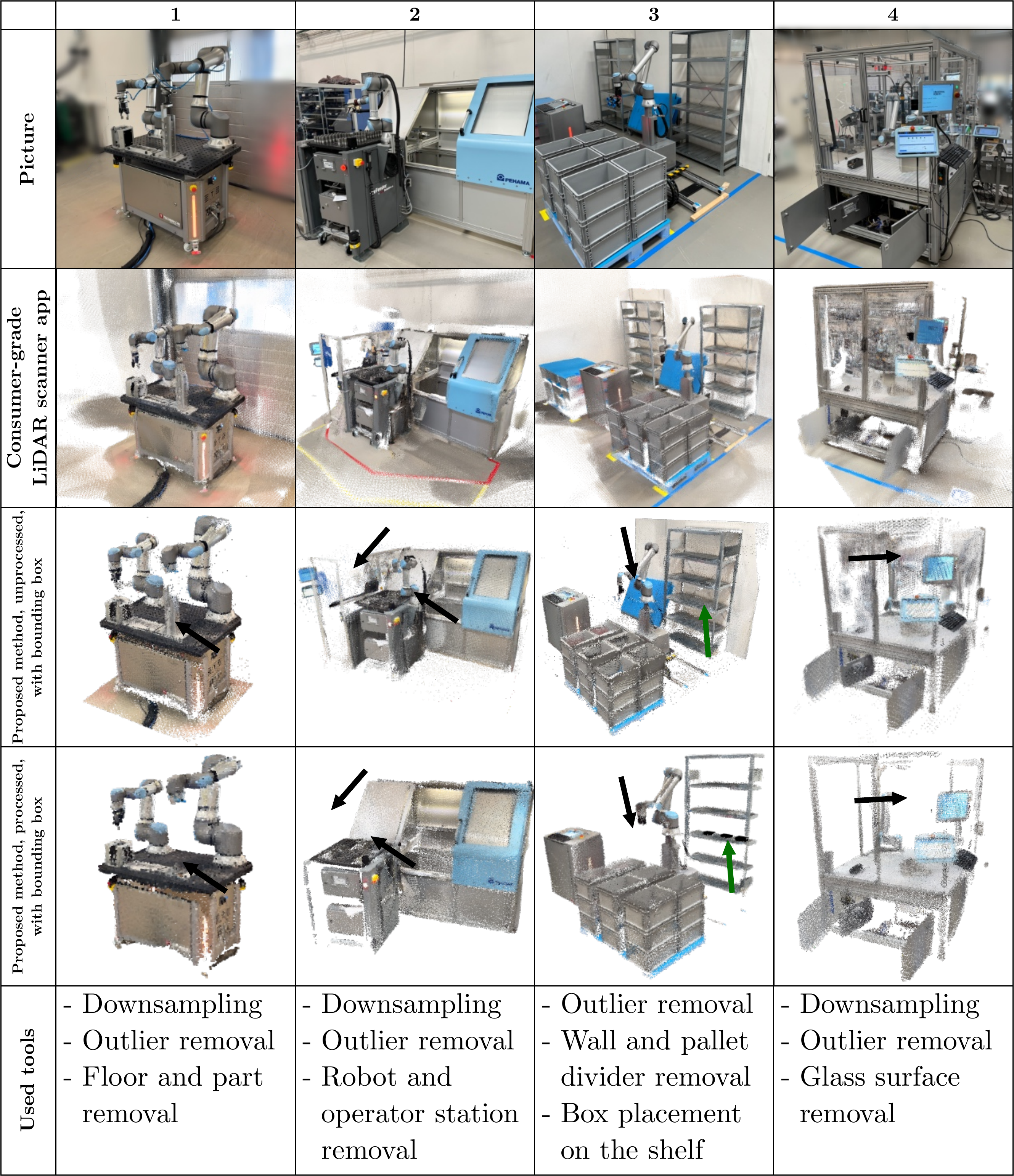}
        \caption{Scans of 4 robot automation workcells utilizing our proposed method:  \ref{workcell:1}) material handling, \ref{workcell:2}) machine tending, \ref{workcell:3}) palletizing, \ref{workcell:4}) assembly. The last row describes the used tools. The arrows present visible changes.}
        \label{fig:WorkcellExamples}
    \end{center}
\end{figure}

To evaluate the usability of the results, we have also designed four robot automation workcells to showcase how our proposed method could be used in real-time and within real-world point cloud processing. All the examples can be seen in \autoref{fig:WorkcellExamples} and are described below:
\begin{enumerate}[wide]
    \item Material handling: this workcell shows a simple dual robot pick-and-place task. Using our toolbox, we have cleaned up the scan and removed the picking source for a scenario where it is replaced by, e.g., a conveyor belt. \label{workcell:1}
    \item Machine tending: this automation uses a robot to load and unload the CNC machine. We used available tools to clean up the scan to remove the operator station that is not needed in the automated workcell, and a robot for a scenario when the reach of the current one is not sufficient and needs to be changed. \label{workcell:2}
    \item Palletizing: the robot is used to un/load pallets of boxes and place/remove them on the shelf. We cleaned up the scan and added virtual boxes on the shelf to show their expected placement. \label{workcell:3}
    \item Assembly: the robot is used for screwing components together. This workcell is isolated behind transparent glass. We used our proposed method to remove incorrectly mapped points in the workcell (on the glass). \label{workcell:4}
\end{enumerate}

\section{Discussion}
\label{sec:discussion}
The benefits of using an AR-based user interface are clearly visible in the presented point cloud captures. For example, it requires much more effort, tools, and time to add a virtual object in the point cloud --- the user must have a point cloud viewer, and CAD modeling software and has to infer a flat surface based on the point representation. Our method, on the other hand, does everything at once --- it acts as a point cloud viewer and CAD modeling software where the pointer is the device's location and gesture input, and flat surfaces are inferred using device AR capabilities. Another example is Workcell \ref{workcell:4}) in \autoref{fig:WorkcellExamples}, where the user can simply use an eraser sponge or brush to remove points that have been incorrectly placed in the 3D space. While the glass is not visible in point cloud representation, the user can easily infer incorrectly placed points by being physically in the space and having freedom of movement.

The main pitfalls of the solution are, even after processing, the inaccuracies of the widely available LiDAR scanners. As it can be seen in \autoref{fig:3Dmodels}b), the legs of the chair are not modeled at all (although, we can add them as primitives later on). Moreover, the mean error is within $1cm$ which is not sufficient for very precise applications, like robot welding. We are, however, mitigating a lot of inaccuracies with user input with the proposed solution. Another problem is that we solely use the device's internal tracking system (visual and inertial odometry) which can accumulate error over time, especially when using an eraser sponge tool that requires moving the device close to a surface we want to erase. When that happens, the tracking system might not work properly because the camera gets less light and there are fewer distinctive features to anchor the virtual representation on. The user might observe drift and erase wrong points. Once, there are enough distinctive features, we realign the virtual and real world.

This paper focuses on improving the quality of point cloud scanning for robot automation applications. This solution can be used for quick scans to perform evaluation of the potential robot deployment. However, with the improved accuracy of the captured workcell, the use cases within robotics expand. For example, with a processed scan, we can evaluate the occupancy map of the workcell by replacing points with voxels (like Octomap) and perform path planning of the trajectory. Furthermore, this can be used for finding the optimal placement of the robot base. It can also be used for coarse robot programming with fine-tuning of the movements by the user.

In the future, more tools can be added based on the research on user needs for point cloud processing within robotics domain. For example, placing of primitives might not be enough to model all the objects, so possibility to add sampled CAD models could be considered.

\section{Conclusions}
\label{sec:conclusions}

Our novel AR-enabled point cloud processing tool streamlines workcell modeling. By capturing and refining point clouds directly on mobile devices, we bridge the gap between rapid capture and precise parametric modeling. We implement a toolbox with five point cloud processing tools: outlier removal, downsampling, point creation using primitives, eraser sponge, and eraser spray. We compare our method with an off-the-shelf point cloud LiDAR scanner app and show the improvements in the accuracy of the obtained capture. With examples of robot automation workcells, we show how this tool can be used for robot deployment purposes.


\section*{ACKNOWLEDGMENT}
\noindent This paper was produced in collaboration with Universal Robots.
This work is partially funded by Innovation Fund Denmark, as part of Industrial PhD program.





\bibliographystyle{IEEEtran}
\bibliography{IEEEabrv,IEEEexample,mybibliography}



\end{document}